\documentclass[a4paper,11pt]{article}
\pdfoutput=1 % if your are submitting a pdflatex (i.e. if you have
\usepackage{jinstpub} % for details on the use of the package, please
\usepackage[utf8]{inputenc}
\usepackage[czech, english]{babel}
\usepackage{cleveref}
\usepackage{import}
\usepackage{xifthen}
\usepackage{pdfpages}
\usepackage{transparent}
\usepackage[subtle]{savetrees}

\graphicspath{{./fig/}}

\title{Track Lab: extensible data acquisition software for fast pixel~detectors, online analysis and automation}

\author[a,b]{P. Mánek,\note{Corresponding author.}}
\author[a,c]{P. Burian,}
\author[a]{E. David-Bosne,}
\author[a]{P. Smolyanskiy}
\author[a]{and B. Bergmann}

\affiliation[a]{Institute of Experimental and Applied Physics, Czech Technical University in Prague,\\Husova 240/5, Prague, 110 00, Czech Republic}
\affiliation[b]{Department of Physics and Astronomy, University College London,\\Gower Street, London, WC1E 6BT, United Kingdom}
\affiliation[c]{Faculty of Electrical Engineering, University of West Bohemia,\\Univerzitní 2795/26, Pilsen, 301 00, Czech Republic}

\emailAdd{petr.manek@utef.cvut.cz}

\abstract{%
  Fast, incremental evolution of physics instrumentation raises the question of efficient software abstraction and transferability of algorithms across similar technologies. This contribution aims to provide an answer by introducing Track Lab, a modern data acquisition program focusing on extensibility and high performance. Shipping with documented~API and more than 20~standard modules, Track Lab allows complex analysis pipelines to be constructed from simple, reusable building blocks. Thanks to multi-threaded infrastructure, data can be clustered, filtered, aggregated and plotted concurrently in real-time. In addition, full hardware support for Timepix2, Timepix3 pixel detectors and embedded photomultiplier systems enables such analysis to be carried out online during data acquisition. Repetitive procedures can be automated with support for motorized stages and X-ray tubes. Freely distributed on 7~popular operating systems and 2~CPU architectures, Track Lab is a versatile tool for high energy physics research.
}


\arxivnumber{2310.08974}

\proceeding{24$^{\text{th}}$ International Workshop on Radiation Imaging Detectors\\
  June 25-29, 2023\\
  Oslo, Norway}

\begin{document}
\maketitle
\flushbottom

\section{Introduction}
\label{sec:intro}

Fast-paced development of physics instrumentation imposes increasing demand on capabilities of data acquisition (DAQ) software. For instance, Timepix4~\cite{llopart2022timepix4} --- the latest generation of Timepix detectors --- comprises $512\times448$~pixels, which is approximately $3.5\times$ that its predecessor Timepix3~\cite{poikela2014timepix3} ($256\times 256$). Furthermore, Timepix4 achieves per-pixel bandwidth of 10.8~kHz, which is nearly $8.3\times$ that of Timepix3 (1.3~kHz). Given the scale of hardware advances, it would appear reasonable to drive Timepix4 with bespoke DAQ software that is fully capable of utilizing its features. On the other hand, Timepix4 is in many ways similar to its predecessors, for which DAQ~toolkits already exist~\cite{Burdaman,broulim2019j,al2019pymepix}. In spite of technological differences between the chips, they undergo similar calibration and diagnostic procedures, use comparable configuration parameters, and their outputs are interpreted in analogous ways. This raises the question of efficient architectural abstraction in software design and the associated trade-off between performance and transferability of algorithms. An ideal solution would deliver both of these qualities.

While there already exist programs that could be viewed as reasonable first choices for this purpose, in practice the range of options is limited by various considerations. Out of readily available academic packages~\cite{turecek2011pixelman,broulim2019j,al2019pymepix,Burdaman}, many are no longer actively maintained. This may shift one's focus towards industrial solutions~\cite{PixetPro} or universal DAQ~frameworks~\cite{kalkman1995labview,gilat2004matlab}. Unfortunately, researchers may find these undesirable for a multitude of reasons. First, many such packages are proprietary, which makes them hard to inspect whenever bugs are suspected. Due to licensing and procurement costs, their adoption must be considered carefully, and their vendors are not well-motivated towards easy integration with other programs. Secondly, existing solutions are usually specific to particular readout systems. This prevents transferability of algorithms across similar technologies and hinders replication of results between systems that do not necessarily share identical infrastructure. Finally, universal DAQ~frameworks are often functionally incomplete. Relying on a generic design in order to appeal to wide user base, many of their popular features tend to be implemented in community repositories with varying levels of reliability, documentation and support.

Aiming to overcome these difficulties, this contribution presents Track Lab --- a modern software package for high-performance DAQ and real-time analysis. Originally designed for controlling pixel detectors, Track Lab has been generalized and adapted for diverse scientific instruments, such as photomultipliers, X-ray tubes and motorized stages. To this end, Track Lab offers modular architecture that can be easily extended to add support for custom hardware or program logic. To achieve high performance, Track Lab relies on industry-standard middleware~\cite{hintjens2013zeromq} and multi-threading that are designed to scale efficiently across many CPU cores (or possibly multiple computers in the future). Offering wide compatibility with popular operating systems and CPU architectures, Track Lab is a free and open-source tool suitable for applications in high energy physics research.

\section{Architecture}
\label{sec:architecture}
\begin{figure}[htbp]
    \centering
    \includegraphics[width=.6\textwidth]{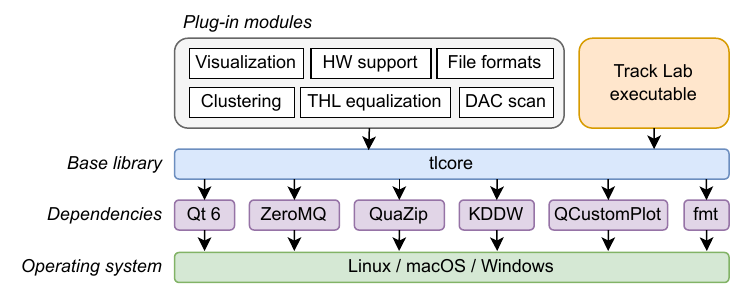}
    \caption{\label{fig:arch-components}Architecture of Track Lab. Boxes correspond to linking units (libraries, executables), arrows indicate dependency relationships. For brevity, only a subset of actual dependencies and modules is displayed.}
\end{figure}

Software architecture of Track Lab was designed according to the following principles:%
\begin{description}
    \item[Extensibility]
    Features are organized in a singular core and a number of independent plug-in modules (illustrated in~\Cref{fig:arch-components}). While the core defines application programming interface (API), modules implement individual features. Users can select modules according to their needs, platform, or choose to extend the software by supplying custom modules.

    \item[Reusability]
    Complex procedures are broken down into single-purpose components, which can be combined like building blocks. Consequently, frequently performed tasks (e.g.,~calibration) are only implemented once at the highest level, abstracted from hardware peculiarities.

    \item[Parallelism]
    Wherever reasonable, performance is optimized by multi-threaded and vectorized data processing. This enables scalability on systems with compatible CPU~hardware, while remaining compatible with less powerful systems thanks to preemptive scheduling.

    \item[Platform-independence]
    Implemented using Qt framework~\cite{Dey_Nibedit2021-06-25}, Track Lab makes no material assumptions about the operating system or CPU architecture. Recursively, all dependent libraries are also platform-agnostic.

    \item[Interoperability]
    Functions of controlled instruments are abstracted as capabilities, defined in the core~API. In addition, data streams and file formats are documented, so as to allow easy integration with other software. The source code is available for inspection upon request.

    \item[Automation]
    Thanks to the signal-slot mechanism of Qt, Track Lab components can easily respond to various events of interest. This facilitates orchestration of equipment, which is particularly desirable for long-running, repetitive measurements.
\end{description}

\subsection{Data flow}
\label{sec:data-flow}
\begin{figure}[htbp]
    \centering
    \includegraphics[width=.9\textwidth]{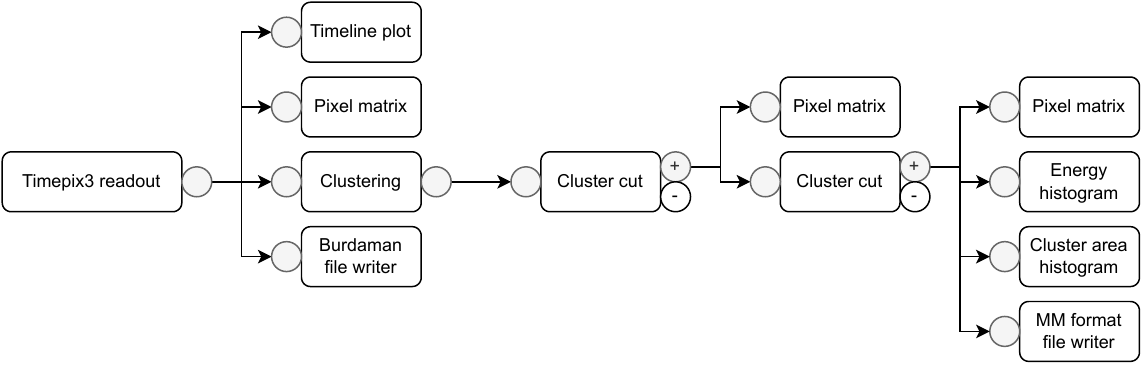}
    \caption{\label{fig:pipeline}Typical topology of a pipeline graph with a single device actor (Timepix3 readout, left) and many actors that modify, visualize and save the data stream. Even though some non-device actors are used multiple times, Track Lab permits this analysis to be efficiently performed in real-time.}
\end{figure}

Data flow in Track Lab is formalized as a directed acyclic graph (DAG), which is called the \emph{pipeline graph}. Nodes of this graph represent \emph{actors} --- instances of program logic that act on data.\footnote{Not to be confused with plug-in modules. Modules contain implementation of program logic, including actors.} Each actor can declare an arbitrary number of inputs and outputs, to which directed edges are connected in order to facilitate data or control flow. In practice, actors usually represent physical devices or data operations (e.g.,~visualization, aggregation, filtering). While it only makes sense for a single device to be used at most once in the graph, other actors can be instantiated arbitrarily many times. Users have complete control of actors as well as connections between them. This way, complex processing topologies can be easily constructed by graphical programming, as illustrated in~\Cref{fig:pipeline}.

Actors are configured from a dedicated screen called the \emph{front panel}, which is populated by panels that correspond to actors in the pipeline graph. For convenience, these panels can be detached, dragged around and docked at various locations within program window to form customized layouts. Together, the topology of the pipeline graph along with the state of the front panel comprise a \emph{session} that can be saved to, or loaded from a persistent file. This saves time by allowing for complex or frequently used topologies to be constructed only once and reused.

\subsection{State machine}
\label{sec:state-machine}
To achieve parallel processing, actors work with data on dedicated CPU threads in the background. This conveniently avoids blocking the user interface when data rates are high, and enables concurrency on systems with compatible hardware. On the other hand, the main drawback of this method is increased synchronization overhead between threads. To minimize this undesirable effect, actors implement a finite state machine (FSM) that synchronizes threads in its state transitions.

\begin{figure}[htbp]
    \centering
    \newlength{\sdheight}
    \setlength{\sdheight}{4.6cm}
    \includegraphics[height=\sdheight]{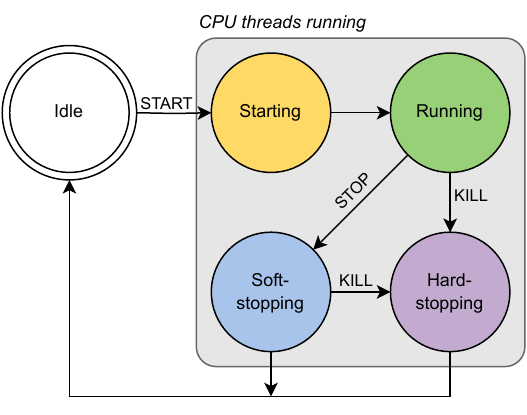}%
    \quad%
    \includegraphics[height=\sdheight]{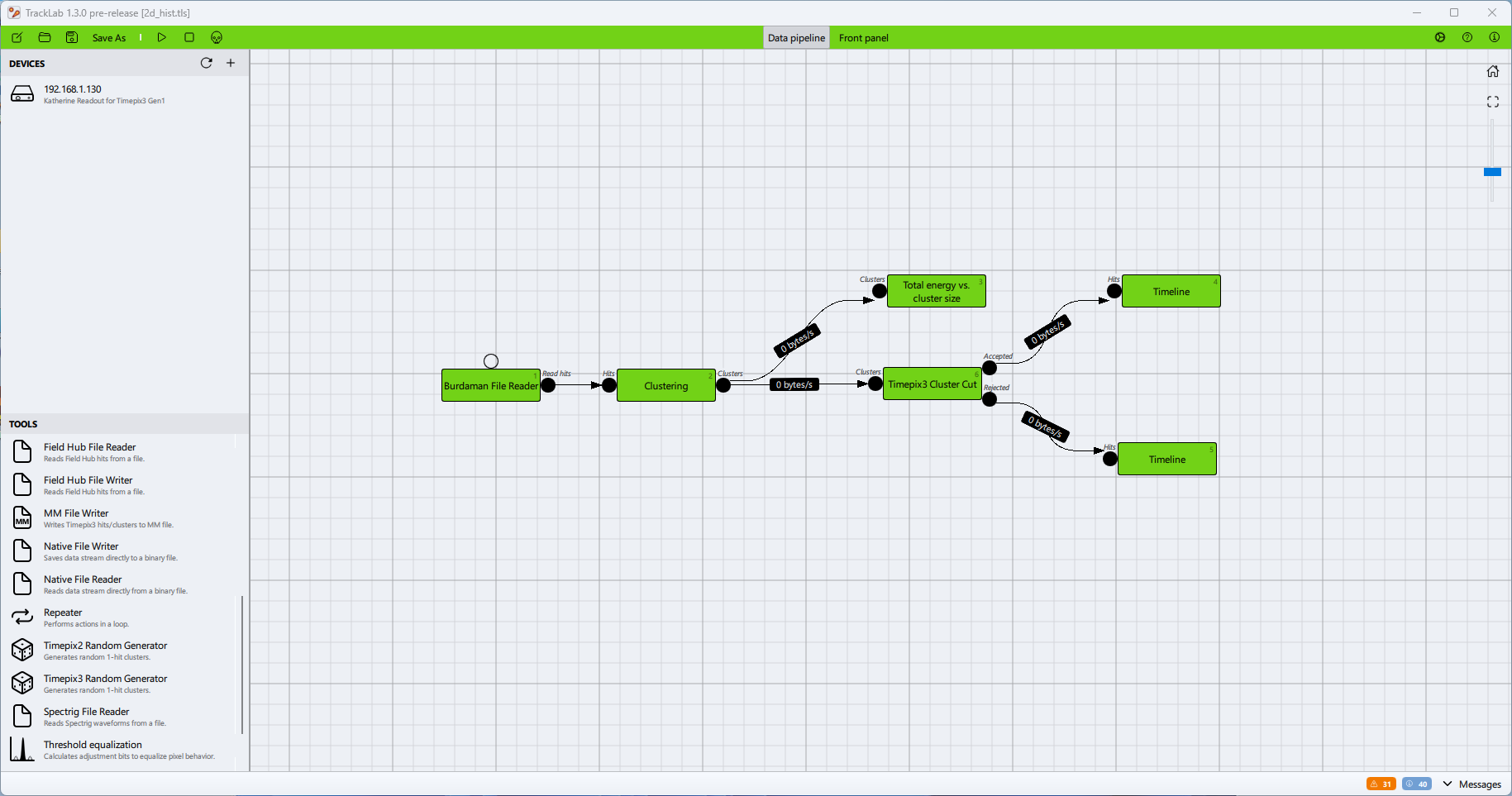}%
    \caption{\label{fig:state-diagram}On the left, state diagram of a single actor. The initial state is marked with double border, labeled state transitions can be initiated by the user. States are color-coded to match the user interface (on the right).}
\end{figure}

At any one time during operation, individual actors are found to be in one of several states (shown in~\Cref{fig:state-diagram}) that describe whether their threads are idle, processing data, or about to transition. In the idle state, actors can be added to or removed from the pipeline graph, and connections between actors can be created. Since no data is flowing, actor configuration on the front panel is disallowed. Conversely, in the running state the pipeline graph is locked out, data flows between actors and the front panel is enabled.

Users can control state of the entire session, which is a combination of the states of all its actors. In addition to conventional ``start'' and ``stop'' actions, which initialize and terminate data flow respectively, the FSM also implements a ``kill'' action. While ``stop'' instructs all actors to return to idle state as soon as all data is processed, ``kill'' interrupts processing immediately at all cost, potentially discarding unconsumed data in the process. This is advantageous in cases where analysis needs to be aborted after large volume of undesirable data is introduced (e.g., by noisy pixels), and their processing would waste computing time or yield results of little value.

\subsection{Performance optimizations}
\label{sec:optimizations}
To increase processing rates, Track Lab implements variety of performance optimizations. These aim to implement lockless synchronization primitives, increase the amount of data processed per single CPU clock cycle (e.g.,~vectorization) and reduce overhead in performance-critical parts of the program (e.g.,~unnecessary copying). In addition, data transfers between actors are facilitated by ZeroMQ~\cite{hintjens2013zeromq}, an asynchronous message-passing middleware. This allows connections between actors to be implemented in a scheme analogous to BSD sockets, which has several advantages:%
\begin{description}
    \item[Transparent multiplexing]
    ZeroMQ efficiently implements a publish-subscribe pattern with one-to-many multiplicity. Consequently, in the pipeline graph a single output can be connected to multiple inputs without significant increase in memory footprint.

    \item[Interoperability]
    Even though data access is currently limited to actors running in Track Lab's main process, in the future the in-process transport protocol can be substituted for a suitable inter-process alternative. This would allow actors to be implemented in any of the 28~programming languages supported by ZeroMQ, massively increasing interoperability with other programs.

    \item[Horizontal scalability]
    By using a transport based on TCP or UDP, data flow between actors can be offloaded to conventional network infrastructure. This would allow subgraphs of the pipeline graph to be distributed to remote machines, increasing overall computing power if desired.
\end{description}

For file reading and writing, Track Lab uses memory-mapped operations, which achieve acceleration by exploiting the paging and virtual addressing mechanism. Similarly to shared dynamically-linked libraries or swapped memory pages, the operating system is instructed to map file segments into memory, where they can be directly accessed as contiguous ranges of bytes. This enables use of conventional optimization strategies (e.g.,~caching, prefetching, dirty pages), which are usually implemented efficiently in the operating system or in hardware. Due to direct memory access, performance is improved as buffers do not have to be copied and exchanged between the software and the operating system. Consequently, as long as access patterns are reasonably predictable, potentially expensive I/O~operations are effectively carried out asynchronously by the operating system without blocking the program from running.

\section{Features}
\label{sec:features}
This section provides an overview of features, which are implemented in modules distributed along with Track Lab core. Thanks to extensible architecture of the software and its documented~API, users may view these as reference implementation or inspiration for development of custom modules.

\subsection{Hardware support}
\label{sec:hardware}
\begin{figure}[htbp]
    \centering
    \newlength{\shheight}
    \setlength{\shheight}{3.05cm}
    \includegraphics[height=0.8\shheight]{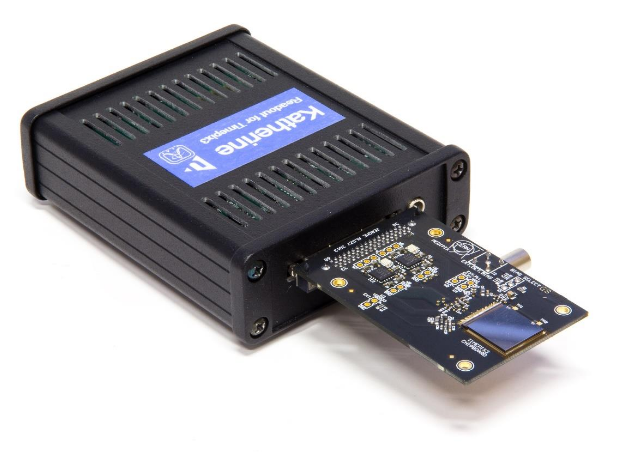}%
    \quad%
    \includegraphics[width=\shheight,angle=90]{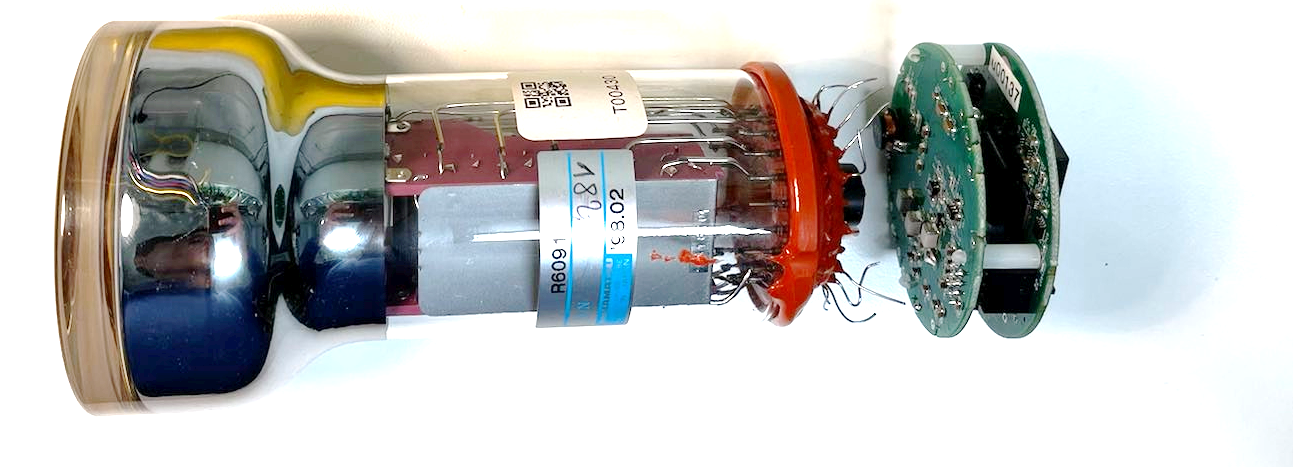}%
    \quad%
    \includegraphics[height=0.7\shheight]{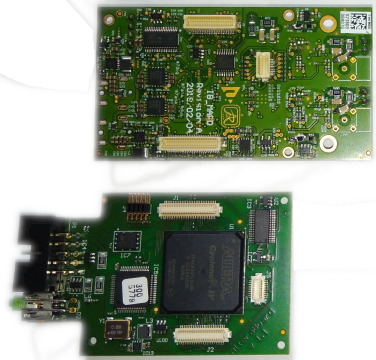}%
    \quad%
    \includegraphics[width=\shheight,angle=90]{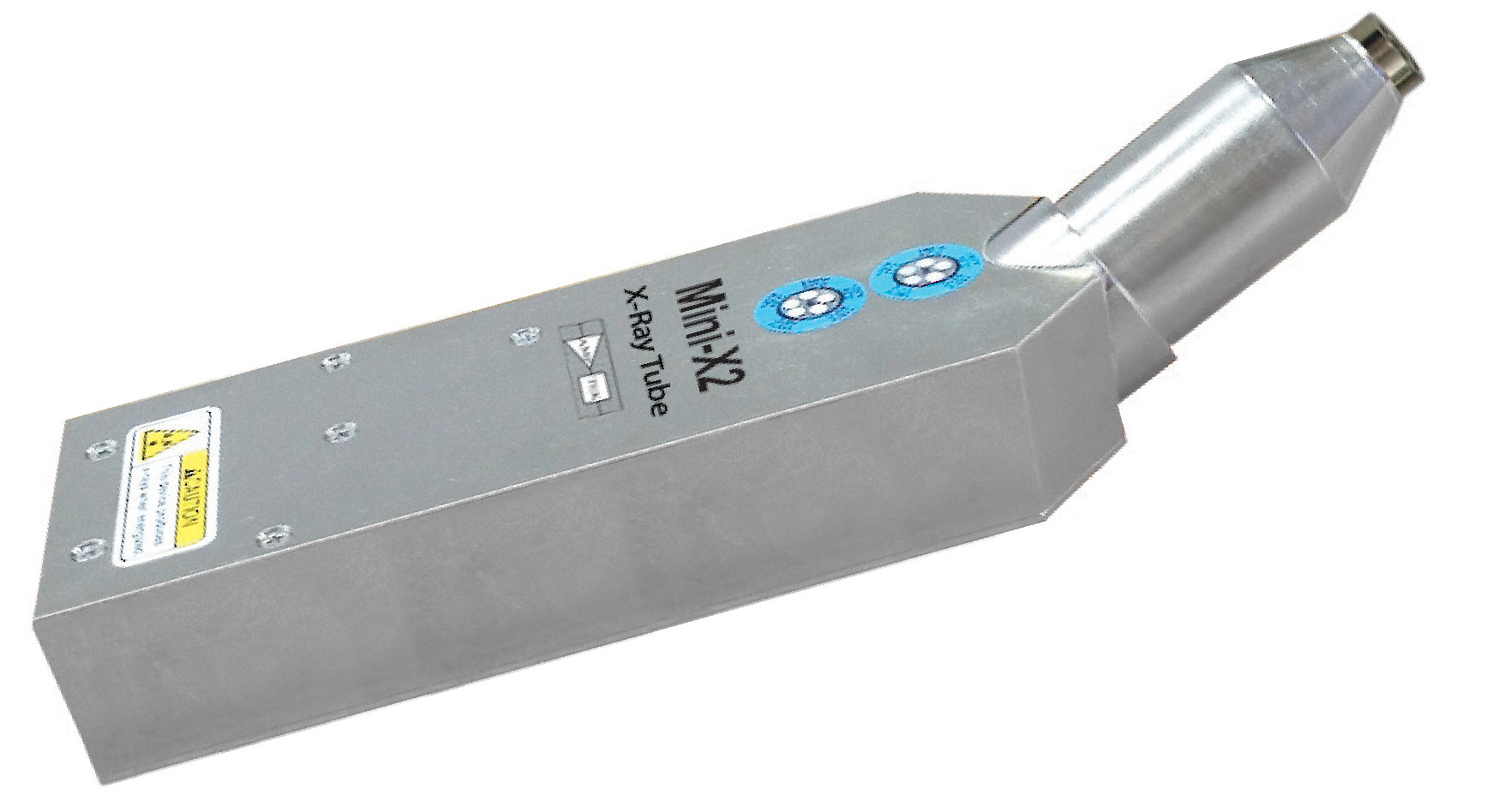}%
    \quad%
    \includegraphics[height=0.6\shheight]{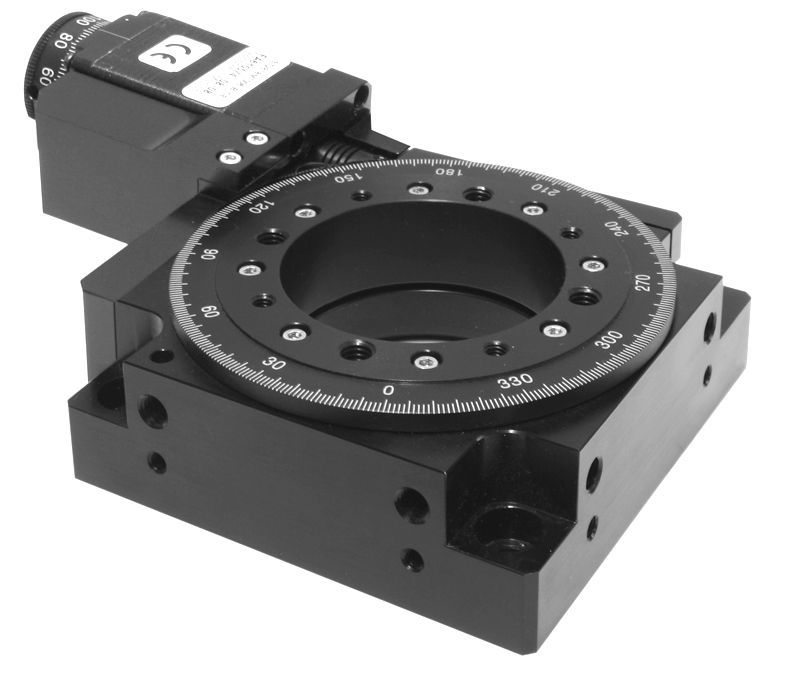}%
    \quad%
    \includegraphics[height=\shheight]{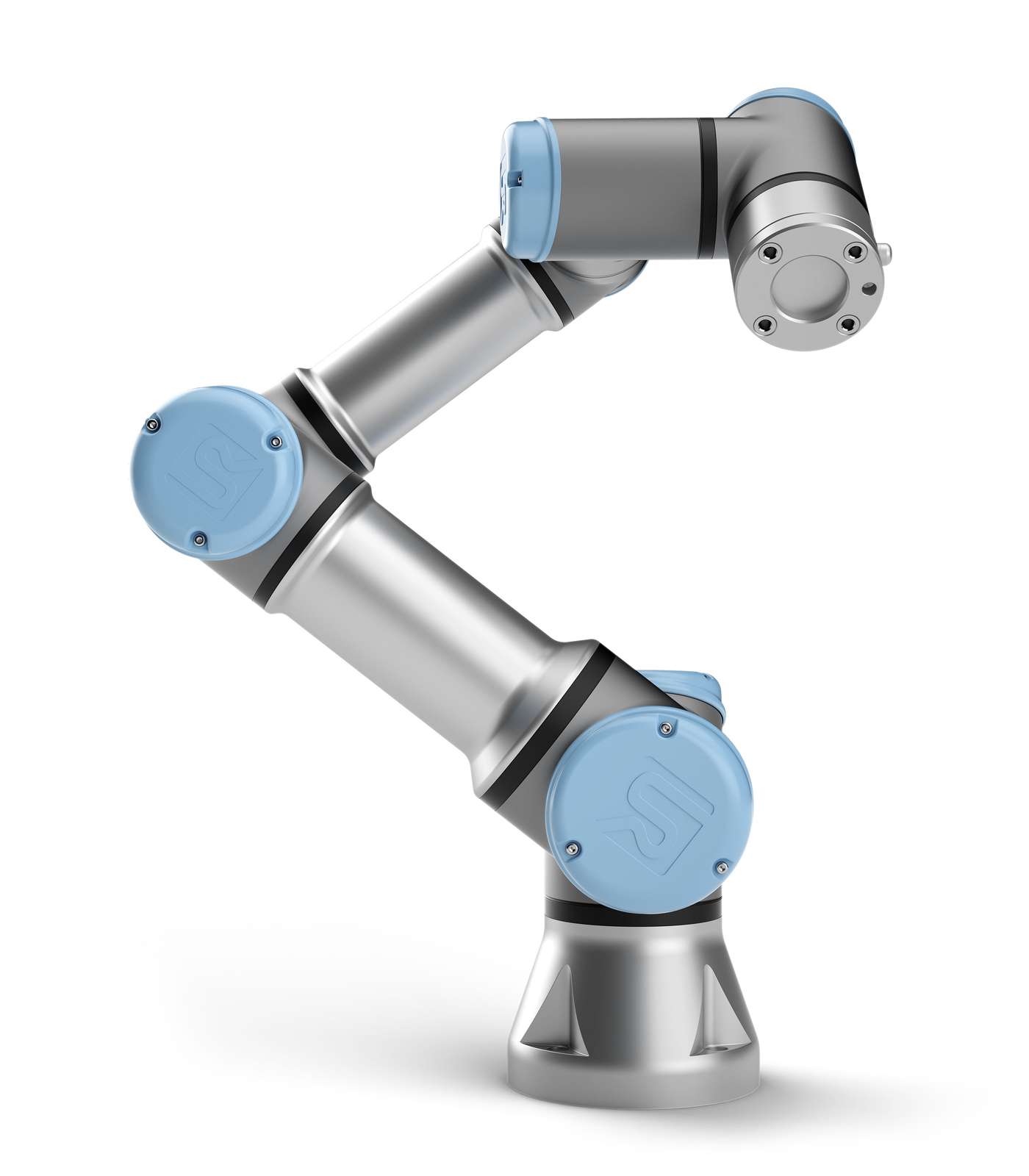}%
    \caption{\label{fig:supported-hardware}Hardware compatible with Track Lab (not to scale). From the left, Katherine with Timepix3, MicroDAQ, Spectrig, Amptek\textsuperscript{\tiny\textregistered}~Mini-X2, Standa~8MR190-2 rotation stage and Universal~Robots UR3e.}
\end{figure}

The following modules are responsible for low-level comunication with hardware, adding the capability to control and retrieve data from a broad range of devices (shown in~\Cref{fig:supported-hardware}).

\begin{description}
\item[Katherine]
This module controls Katherine readouts~\cite{burian2017katherine,burian2020ethernet} for Timepix3~\cite{poikela2014timepix3} and Timepix2~\cite{wong2020introducing} over UDP, USB or PCI-e. In addition to simultaneously taking data from up to 8~chips, it can configure the on-board bias power supply, signal generator and GPIO pin functions.

\item[MicroDAQ]
This module drives arrays of 28~MicroDAQ readouts~\cite{kauer2019scintillator} for Hamamatsu and HZC photomultiplier tubes (PMT) over TCP. It can display and configure high voltage of the PMT as well as run data acquisitions at the rate of approximately 29~kHz/channel.

\item[Spectrig]
This module controls the Spectrig waveform digitizer~\cite{holik2020miniaturized} over USB. It can configure and perform acquisition of wavelets from silicon photomultipliers (SiPM) or other signal sources.

\item[Amptek\textsuperscript{\tiny\textregistered} X-ray tubes]
This module drives the Mini-X2 tube~\cite{MiniX2} over USB. It can display interlock state, configure tube's high voltage, current, and energize the tube.

\item[Standa motors]
This module controls translation and rotation stages driven by the 8SMC5-USB controller~\cite{StandaController}. For each axis, the position can be displayed or modified in steps or degrees.

\item[Universal Robots positioning arm]
This module controls the UR3e multi-axis arm~\cite{UniversalRobotsUR3e} over TCP. It can display and command the position of individual joints, or evaluate inverse kinematic transform and move the arm into a desired pose.
\end{description}

\subsection{Real-time analysis}
\label{sec:analysis}

\begin{description}
\item[Clustering]
This module aggregates pixel hits into clusters (or tracks) by their spatial and temporal locality~\cite{meduna2019real,Manek2018_CUNI}. For each cluster constructed this way, up to 12 attributes can be calculated. Reference counting ensures that only attributes, which are used downstream, are evaluated.

\item[Coincidence finding]
This module filters hits and clusters that coincide with timestamped events of interest (i.e.,~signals on GPIO pins) within adjustable, possibly asymmetric time windows.

\item[Attribute cuts]
This module filters clusters by various attributes (e.g.,~energy, number of pixels, centroids, morphology~\cite{holy2008pattern}). Accepted and rejected clusters can be processed independently.

\item[Time-over-Threshold calibration]
This module evaluates previously fitted calibration function for Time-over-Threshold~\cite{jakubek2011precise}, which allows deposited energy to be calculated in absolute units.

\item[Time-walk correction]
This module evaluates previously fitted function that compensates against the time-walk effect~\cite{turecek2016usb}, responsible for Time-of-Arrival delay in analog signals with larger amplitudes due to constant rise time.
\end{description}

\subsection{Visualization}
\label{sec:visualization}
\begin{figure}[htbp]
    \centering
    \newlength{\vsheight}
    \setlength{\vsheight}{3.7cm}
    \includegraphics[height=\vsheight]{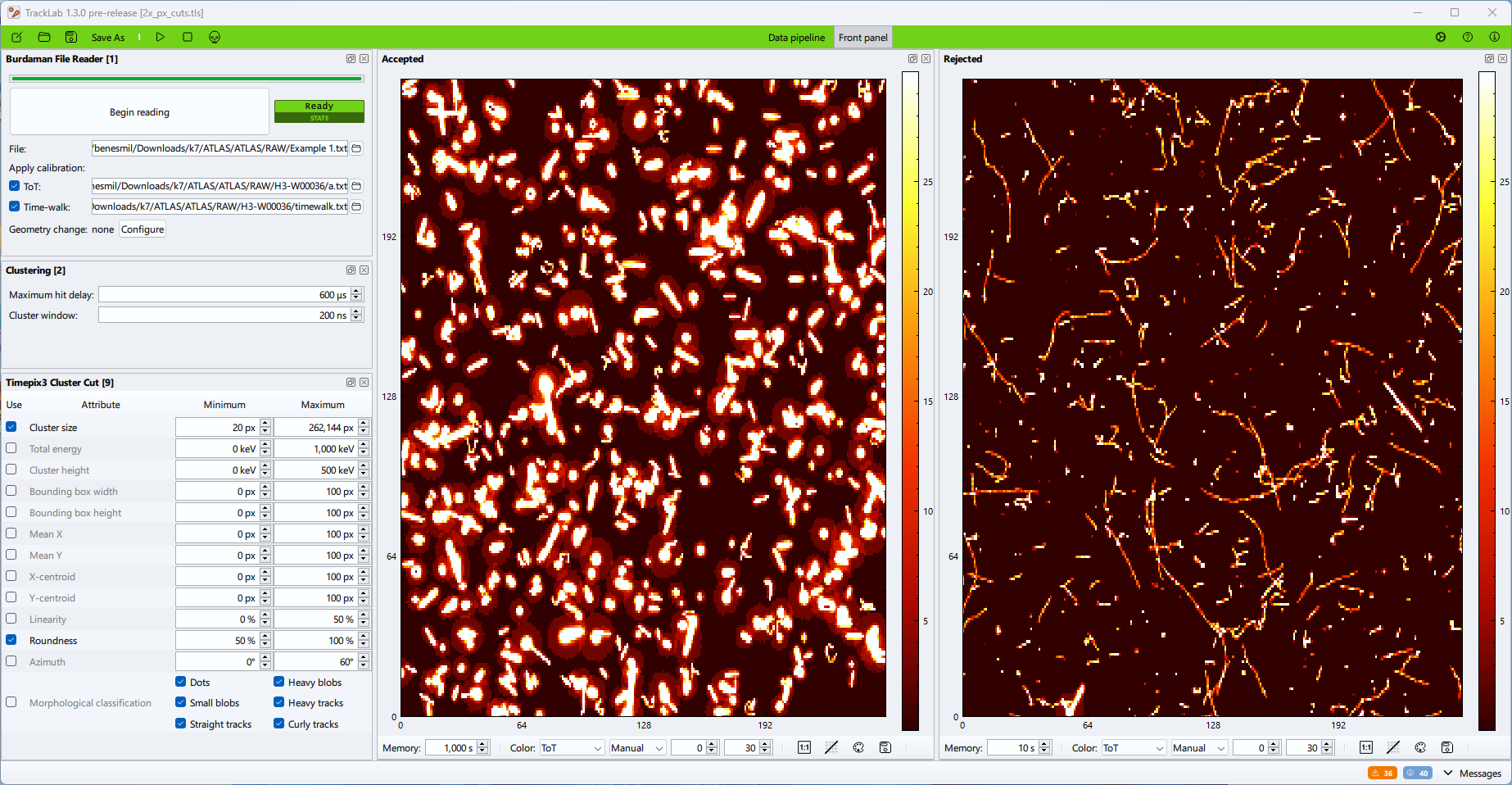}%
    \quad%
    \includegraphics[height=\vsheight]{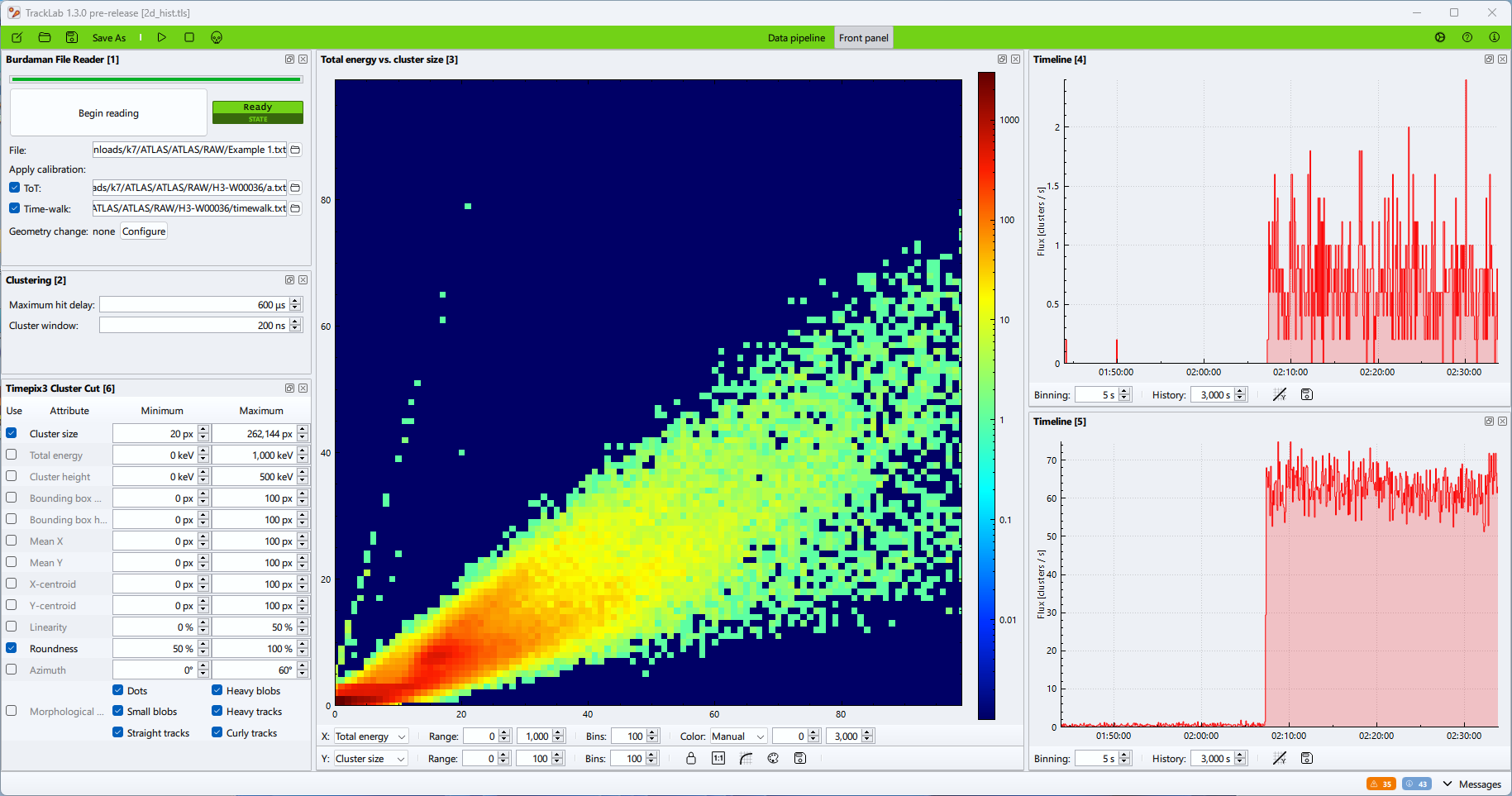}%
    \caption{\label{fig:visualization-screens}Visualization modules displaying data taken with Timepix3 detectors at the ATLAS experiment. In the left window, clusters filtered by size and roundness are fed into two pixel matrices (one for accepted, other for rejected data). In the right window, two-dimensional histogram of cluster energy and cluster size is plotted next to a series of timeline plots.}
\end{figure}

Plotting modules render programmable charts in real-time. For that reason, they are indispensable tools that allow users to visualize and validate results of analysis while data acquisition is still ongoing (as shown in \Cref{fig:visualization-screens}). In all modules listed below, ranges, binning and zoom factors are easily adjustable. Furthermore, the resulting plots can be switched between linear and logarithmic scale and exported to files (bitmap, vector graphics or text files).

\begin{description}
\item[Timeline]
This module displays normalized hit rate or cluster rate as a function of time.

\item[Histograms]
This module plots one- or two-dimensional histograms of selected cluster attributes.

\item[Pixel matrix]
This module displays pixel hits recorded over a tunable period of time. Depending on chip type and used acquisition mode, the color axis can show Time-over-Threshold (optionally with energy calibration), Time-of-Arrival, event counts or other values.
\end{description}

\subsection{Commissioning}
\label{sec:automation}
\begin{figure}[htbp]
    \centering
    \newlength{\asheight}
    \setlength{\asheight}{3.7cm}
    \includegraphics[height=\asheight]{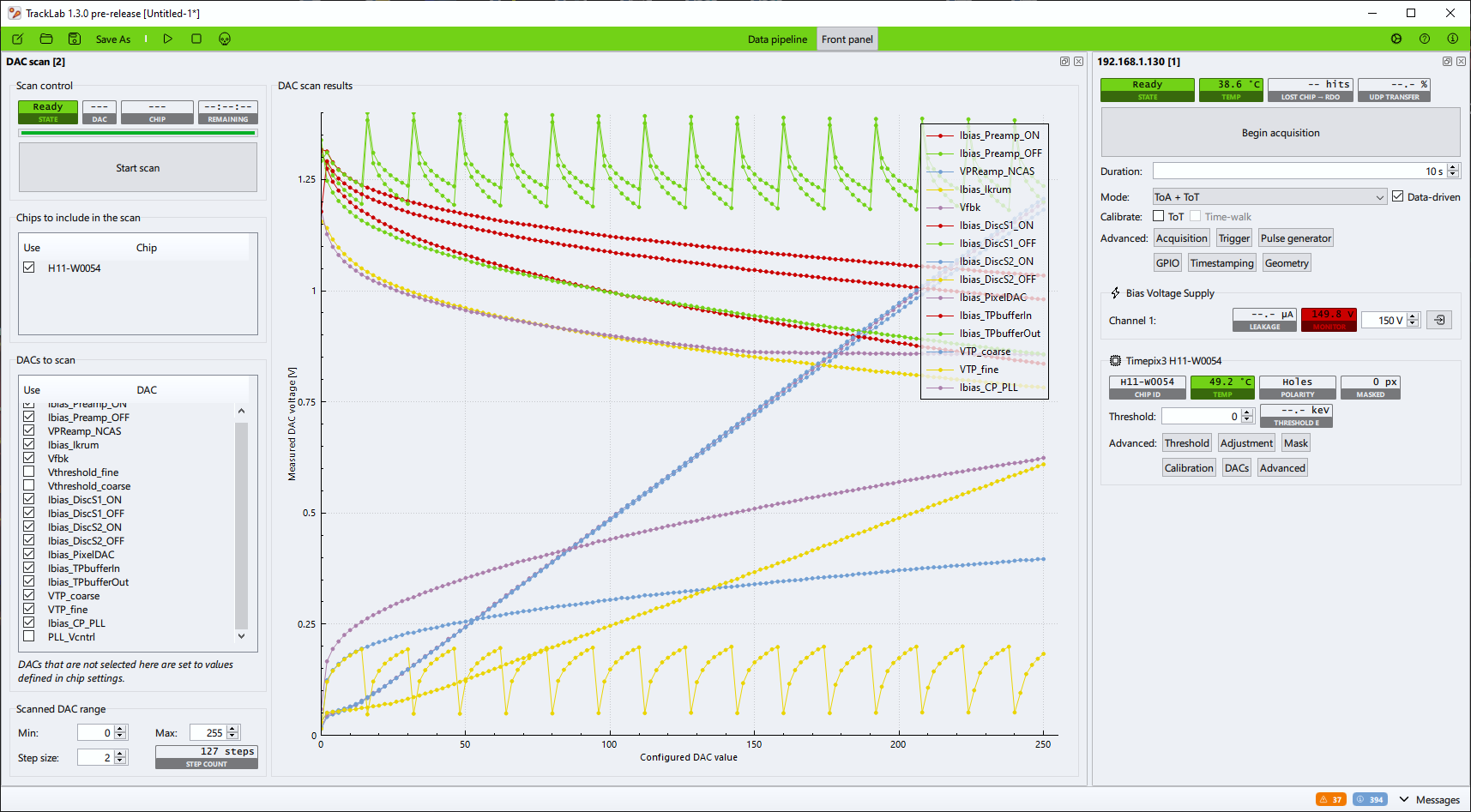}%
    \quad%
    \includegraphics[height=\asheight]{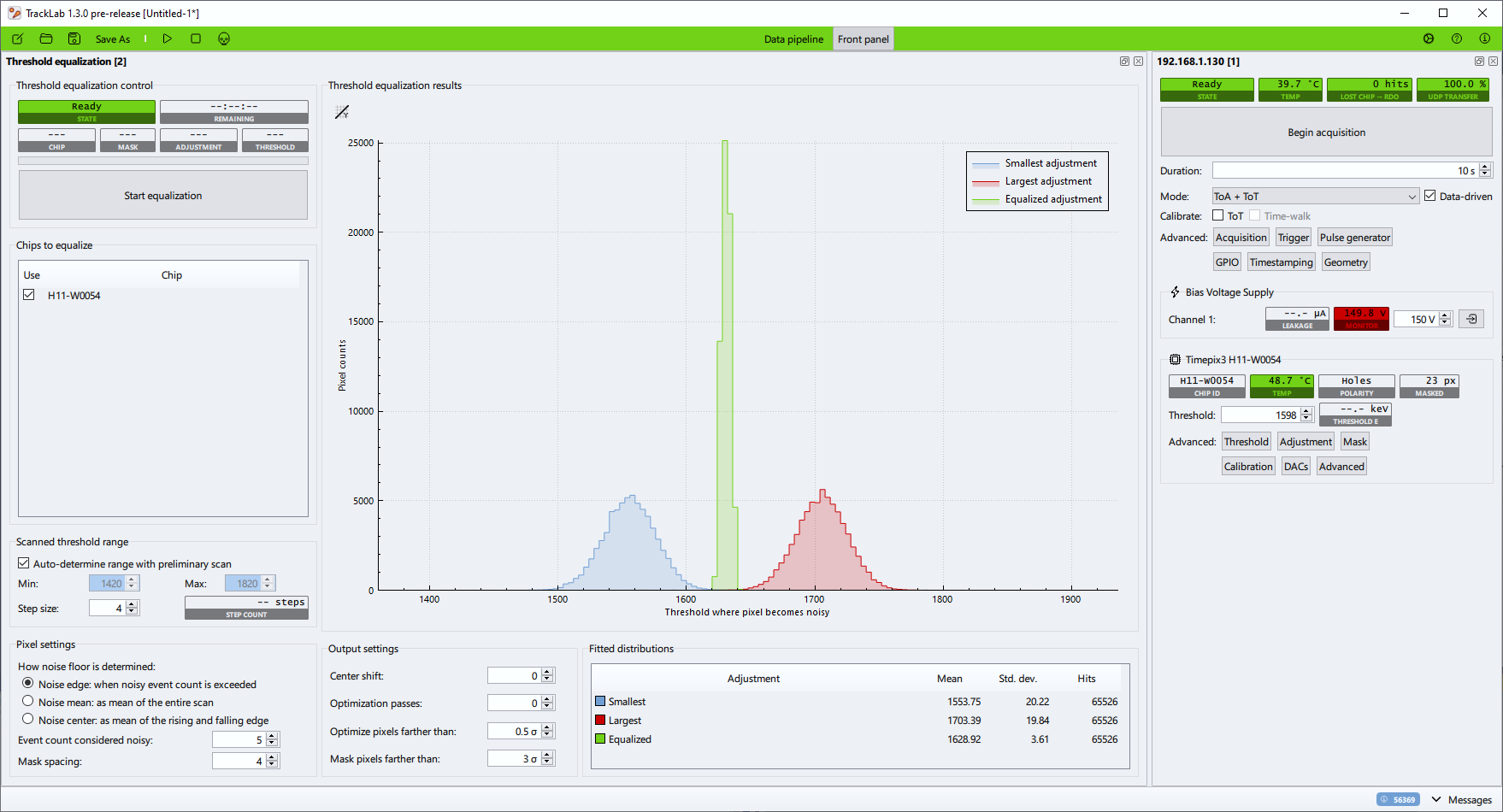}%
    \caption{\label{fig:automation-screens}Select modules in use with Timepix3. In the left window, the DAC scan is displaying effects of varying DAC settings on corresponding generated voltages. In the right window, the threshold equalization module plots pixel noise floor distributions prior to (blue, red) and after (green) equalization.}
\end{figure}

\begin{description}
\item[DAC scan]
This module scans the range of pixel detector digital-to-analog converters (DAC), probes the resulting voltages and plots them as a function of configured values (illustrated in \Cref{fig:automation-screens}).

\item[Threshold scan]
This module scans the threshold range of a pixel detector. If the scan is performed in presence of a mono-energetic radiation source, the resulting response can be fitted to calibrate arbitrary threshold units with well-known energies.

\item[Threshold equalization]
This module determines values of per-pixel adjustment registers, in order to equalize pixel detector response and compensate against manufacturing irregularities. The procedure supports noise-edge, noise-mean and noise-center mode of determining noise floor.
\end{description}

\section{Applications}
\label{sec:applications}
This section provides examples demonstrating practical use of Track Lab in recent research projects.

\subsection{Beam tests for Penetrating Particle Analyzer}
\label{sec:pan-beam-tests}
\begin{figure}[htbp]
    \centering
    \includegraphics[height=3.7cm]{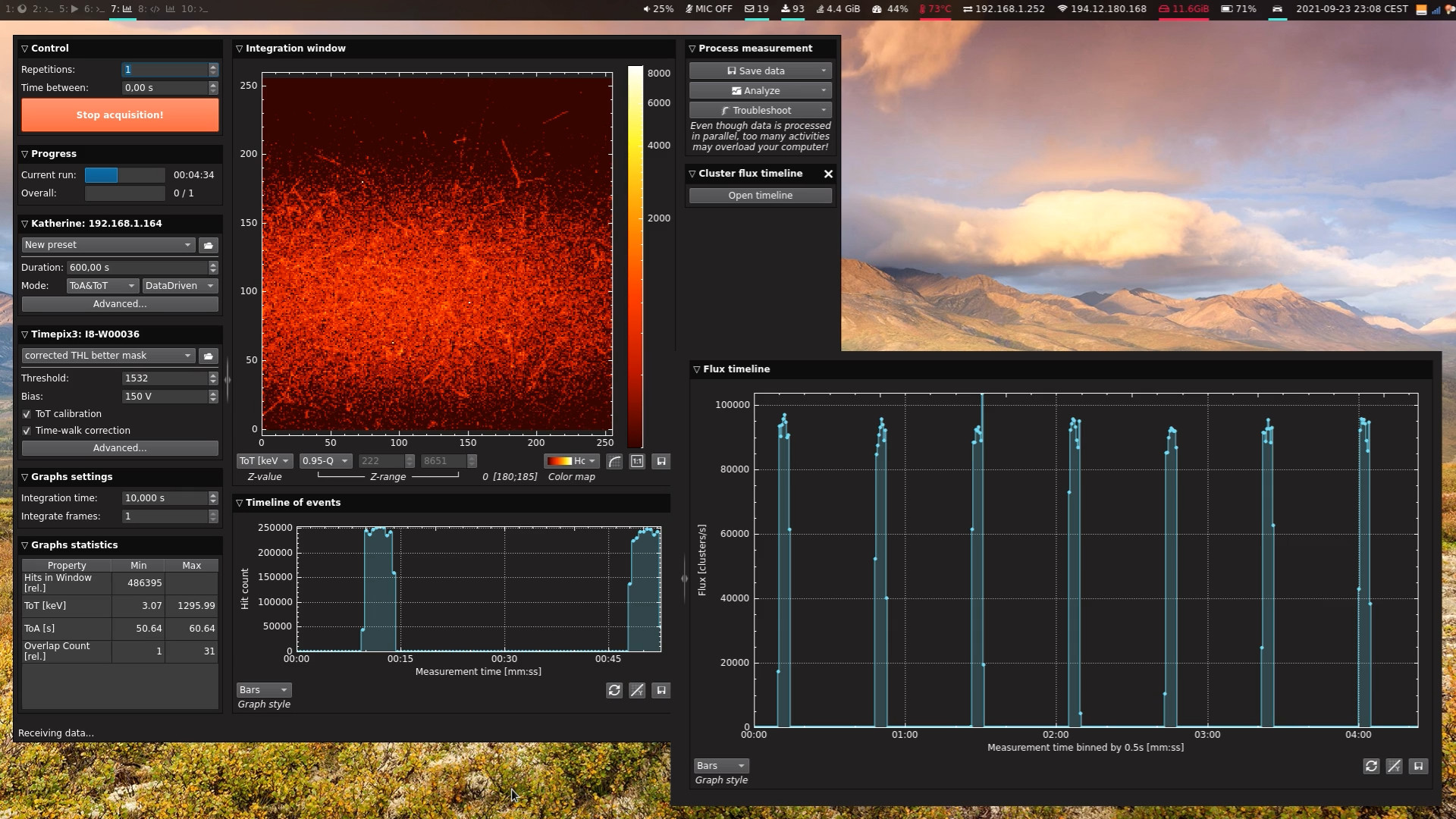}
    \quad%
    \def\svgwidth{6.8cm}
    %% Creator: Inkscape 1.3 (0e150ed6c4, 2023-07-21), www.inkscape.org
%% Accompanies image file 'pan2.mod.annotated.pdf' (pdf, eps, ps)
\begingroup%
  \makeatletter%
  \providecommand\color[2][]{%
    \errmessage{(Inkscape) Color is used for the text in Inkscape, but the package 'color.sty' is not loaded}%
    \renewcommand\color[2][]{}%
  }%
  \providecommand\transparent[1]{%
    \errmessage{(Inkscape) Transparency is used (non-zero) for the text in Inkscape, but the package 'transparent.sty' is not loaded}%
    \renewcommand\transparent[1]{}%
  }%
  \providecommand\rotatebox[2]{#2}%
  \newcommand*\fsize{\dimexpr\f@size pt\relax}%
  \newcommand*\lineheight[1]{\fontsize{\fsize}{#1\fsize}\selectfont}%
  \ifx\svgwidth\undefined%
    \setlength{\unitlength}{532.5bp}%
    \ifx\svgscale\undefined%
      \relax%
    \else%
      \setlength{\unitlength}{\unitlength * \real{\svgscale}}%
    \fi%
  \else%
    \setlength{\unitlength}{\svgwidth}%
  \fi%
  \global\let\svgwidth\undefined%
  \global\let\svgscale\undefined%
  \makeatother%
  \begin{picture}(1,0.54084507)%
    \lineheight{1}%
    \setlength\tabcolsep{0pt}%
    \put(0,0){\includegraphics[width=\unitlength,page=1]{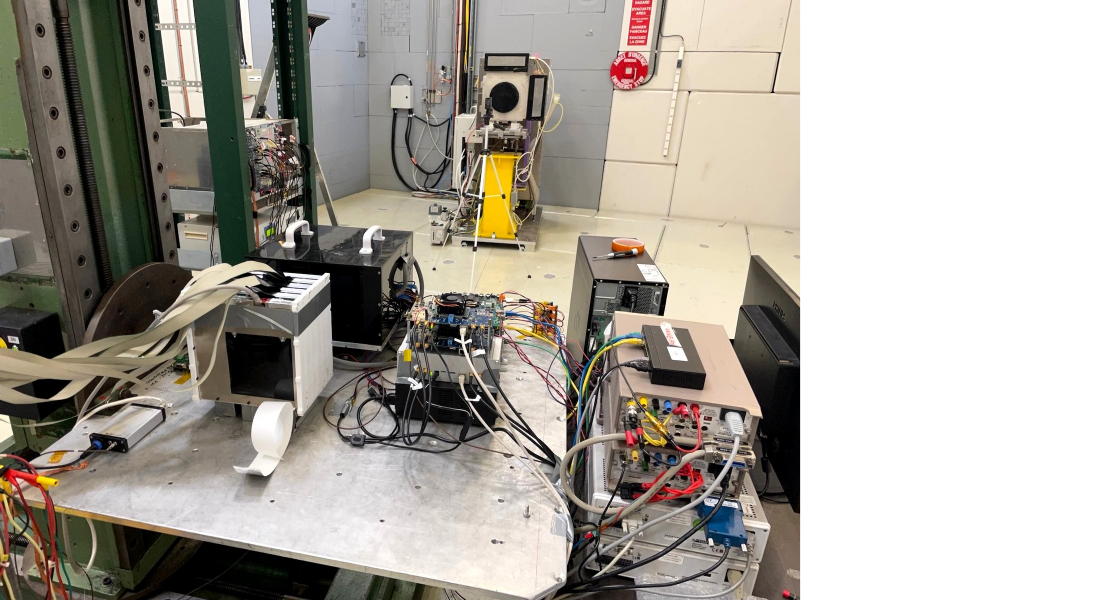}}%
    \put(0.79162331,0.28655067){\color[rgb]{0,0,0}\makebox(0,0)[lt]{\lineheight{1.25}\smash{\begin{tabular}[t]{l}8$\times$ Timepix3\end{tabular}}}}%
    \put(0,0){\includegraphics[width=\unitlength,page=2]{pan2.mod.annotated.pdf}}%
    \put(0.79162331,0.37387461){\color[rgb]{0,0,0}\makebox(0,0)[lt]{\lineheight{1.25}\smash{\begin{tabular}[t]{l}Beam\end{tabular}}}}%
    \put(0,0){\includegraphics[width=\unitlength,page=3]{pan2.mod.annotated.pdf}}%
    \put(0.79162331,0.21049433){\color[rgb]{0,0,0}\makebox(0,0)[lt]{\lineheight{1.25}\smash{\begin{tabular}[t]{l}2$\times$ Katherine\end{tabular}}}}%
    \put(0,0){\includegraphics[width=\unitlength,page=4]{pan2.mod.annotated.pdf}}%
  \end{picture}%
\endgroup%

    \caption{\label{fig:pan-setup}Setup used for a PAN~beam test. Track Lab (left, older version) saves data to files and plots particle flux of the pulsed beam in real-time. On the right, annotated experimental hardware.}
\end{figure}

During development of the Penetrating Particle Analyzer~(PAN)~\cite{wu2019penetrating}, a number of beam tests were carried out at Proton Synchrotron and Super Proton Synchrotron in Geneva, Switzerland. In the course of these tests, 2~Timepix3 Quad detectors (8~chips total) were exposed to a pulsed hadron beam (120~GeV/c, 90\%~pions), as indicated in \Cref{fig:pan-setup}. Detector orientation with respect to the beam was controlled remotely by a Standa rotation stage~\cite{StandaController}. The test procedure involved repeated acquisitions with varying incidence angles, power settings and bias voltages.

Track Lab's real-time visualization capability proved advantageous in initial beam alignment, as it allowed cluster centroids to be plotted in a two-dimensional histogram with sub-pixel resolution. This provided instant feedback for operators. Throughout data taking, automation implemented in the software led to nearly unattended operation, which saved considerable amount of time during repetitive acquisitions with motorized rotation stages. In addition to saving both raw and clustered data, the software was able to display live plots useful for quality control (e.g.,~spectra, fluxes).

\subsection{Contaminated tissue scanning}
\label{sec:tissue-scanning}
\begin{figure}[htbp]
    \centering
    \includegraphics[height=4.81cm]{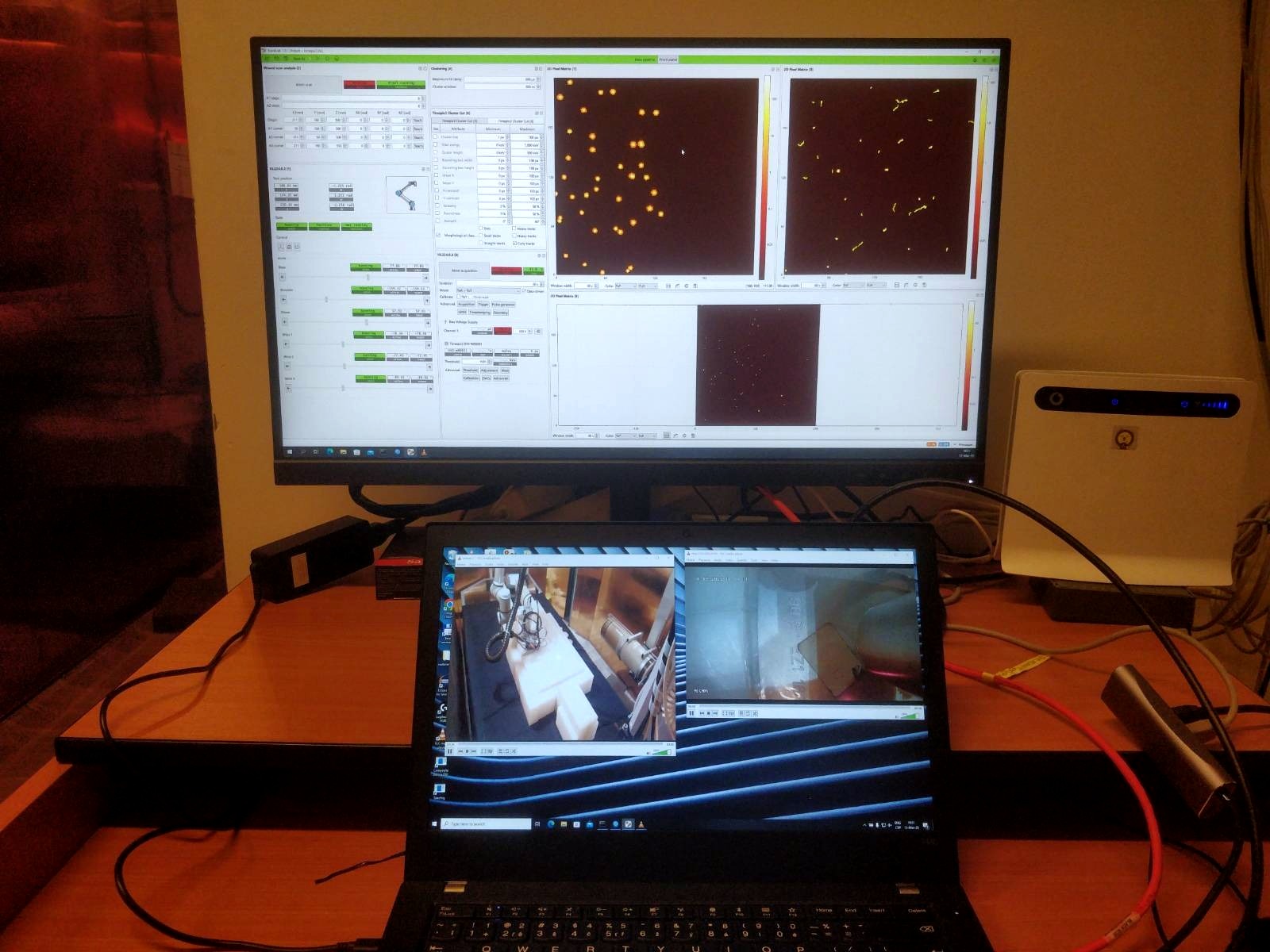}%
    \quad%
    \def\svgwidth{6cm}
    %% Creator: Inkscape 1.3 (0e150ed6c4, 2023-07-21), www.inkscape.org
%% Accompanies image file 'suro2.mod.annotated.pdf' (pdf, eps, ps)
\begingroup%
  \makeatletter%
  \providecommand\color[2][]{%
    \errmessage{(Inkscape) Color is used for the text in Inkscape, but the package 'color.sty' is not loaded}%
    \renewcommand\color[2][]{}%
  }%
  \providecommand\transparent[1]{%
    \errmessage{(Inkscape) Transparency is used (non-zero) for the text in Inkscape, but the package 'transparent.sty' is not loaded}%
    \renewcommand\transparent[1]{}%
  }%
  \providecommand\rotatebox[2]{#2}%
  \newcommand*\fsize{\dimexpr\f@size pt\relax}%
  \newcommand*\lineheight[1]{\fontsize{\fsize}{#1\fsize}\selectfont}%
  \ifx\svgwidth\undefined%
    \setlength{\unitlength}{485.15963745bp}%
    \ifx\svgscale\undefined%
      \relax%
    \else%
      \setlength{\unitlength}{\unitlength * \real{\svgscale}}%
    \fi%
  \else%
    \setlength{\unitlength}{\svgwidth}%
  \fi%
  \global\let\svgwidth\undefined%
  \global\let\svgscale\undefined%
  \makeatother%
  \begin{picture}(1,0.79149206)%
    \lineheight{1}%
    \setlength\tabcolsep{0pt}%
    \put(0,0){\includegraphics[width=\unitlength,page=1]{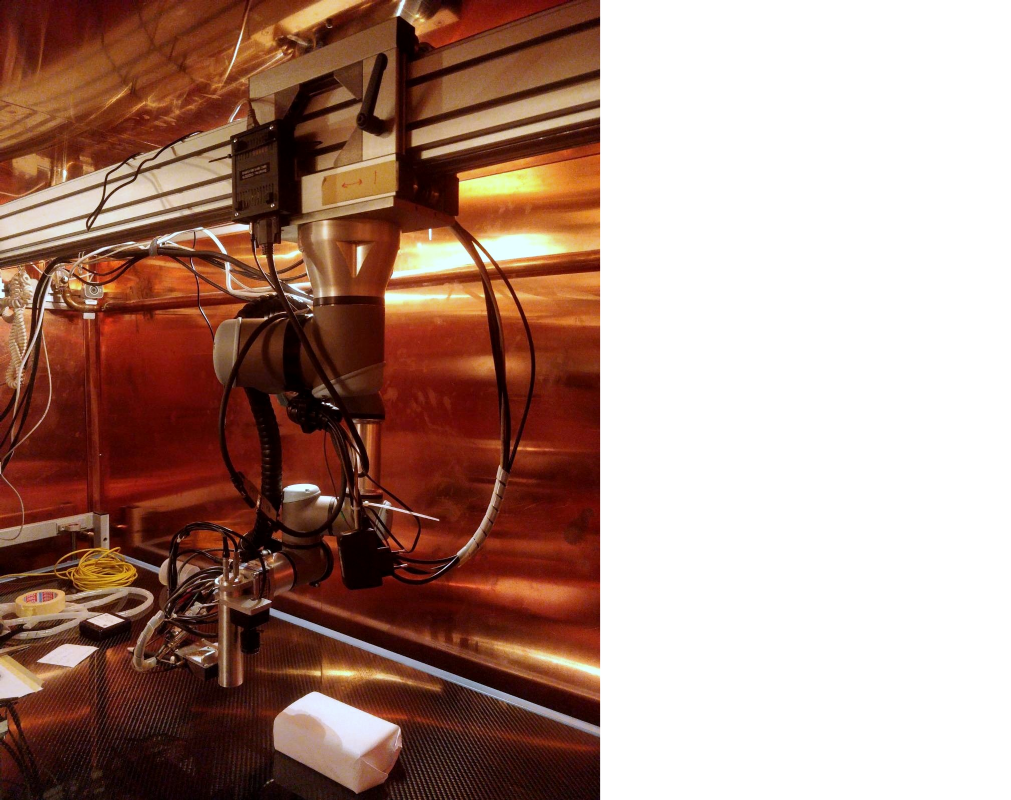}}%
    \put(0.67279813,0.59482893){\color[rgb]{0,0,0}\makebox(0,0)[lt]{\lineheight{1.25}\smash{\begin{tabular}[t]{l}Katherine\end{tabular}}}}%
    \put(0.67349257,0.4212927){\color[rgb]{0,0,0}\makebox(0,0)[lt]{\lineheight{1.25}\smash{\begin{tabular}[t]{l}UR3e arm\end{tabular}}}}%
    \put(0.67195273,0.17473143){\color[rgb]{0,0,0}\makebox(0,0)[lt]{\lineheight{1.25}\smash{\begin{tabular}[t]{l}Timepix3\end{tabular}}}}%
    \put(0.66995999,0.10927564){\color[rgb]{0,0,0}\makebox(0,0)[lt]{\lineheight{1.25}\smash{\begin{tabular}[t]{l}Spectrometer\end{tabular}}}}%
    \put(0,0){\includegraphics[width=\unitlength,page=2]{suro2.mod.annotated.pdf}}%
  \end{picture}%
\endgroup%

    \caption{\label{fig:suro-setup}Setup used for tissue scanning. Track Lab (left) drives robotic arm and filters clusters in order to separate alpha component of the mixed radiation field. On the right, annotated experimental hardware.}
\end{figure}

In collaboration with the Czech National Radiation Protection Institute~(SÚRO), a remote-controlled instrument was developed to analyze biological tissues contaminated with ionizing radiation. The machine (shown in \Cref{fig:suro-setup}) was based on the UR3e~positioning arm~\cite{UniversalRobotsUR3e} with a tool assembly comprising a single Timepix3 chip with Katherine readout and a portable gamma spectrometer with Spectrig digitizer.

In this project, Track Lab was used to control motion of the positioner as well as orchestrate data acquisition. A typical scanning procedure began by teaching the software boundaries of the scanned space (e.g.,~area around a human limb), which would then be systematically examined in adjustable grid pattern. At each point, data was fed into analysis that estimated composition and intensity of the radiation field.

\subsection{Detector networks at ATLAS, MoEDAL}
\label{sec:detector-nets}
Over the last shutdown period (LS2), 2~networks of Timepix3 detectors have been installed at the Large Hadron Collider. At the ATLAS experiment~\cite{burian2018timepix3}, the aim is to characterize mixed radiation fields and monitor luminosity. At the MoEDAL experiment, LHCb~\cite{bergmann2021sissa}, the objective is to reconstruct particle trajectories inside the Vertex Locator (VELO) region, and search for potential magnetic monopole candidates. In both locations, detectors are regularly exposed to harsh radiation conditions, leading to relatively large data rates with considerable potential for radiation damage (i.e.,~noisy or dead pixels). Furthermore, beam operations necessitate that DAQ~infrastructure is capable of absorbing several-hour-long bursts of sustained high-flux data, irrespective of trigger.

To ensure timely communication with a large number of detectors, Track Lab was upgraded with PCI-e communication interface. Compared to a conventional Ethernet connection to Katherine, this link serves the same purpose but offers increased bandwidth. Furthermore, to cope with data bursts the ZeroMQ message-passing middleware was tuned to allow extensive buffering of received data in-memory. Assuming that high-flux conditions can only be sustained for limited time, buffered data can be evacuated from memory during breaks. For increased reliability, tracking of DAQ backpressure was implemented, which can identify bottlenecks in the pipeline graph, and automatically respond when high water marks are exceeded.
\section{Distribution}
\label{sec:distribution}
The latest version of Track Lab supports three widely used desktop operating systems: GNU/Linux, Apple macOS and Microsoft Windows. Linux binaries are provided for the \texttt{x86\_64} and \texttt{aarch64} CPU instruction sets. Users can choose to install packages targeting select distributions (Debian, Enterprise Linux, Fedora, openSUSE, Ubuntu) or use a generic AppImage~\cite{AppImage} that only requires glibc ABI~v2.35 or newer. For macOS, Track Lab is distributed as a Mach-O Universal~2 binary~\cite{Universal2} containing both Intel and M1 machine code. macOS Monterey (v12) or newer is required. Finally, Windows users can choose between a portable software archive, or a one-click Microsoft Installer package. The latest version requires a 64-bit system running Windows 10 or newer.

Track Lab binaries, core API, documentation and other resources are freely available for non-commercial use online\footnote{\url{https://software.utef.cvut.cz/tracklab}} under the CC~BY-NC-SA license~\cite{CreativeCommons}. Access to source code is free and can be arranged upon request, software contributions are welcome.

\section{Conclusion}
\label{sec:conclusion}
This article has introduced Track Lab, a modern DAQ~software directed at applications in physics research and data analysis. The program is based on the principle that complex tasks can be divided into simpler building blocks, and that frequently used procedures can be abstracted and reused in similar hardware without significant performance penalty. To this end, Track Lab provides extensible architecture that ships over 20~modules as well as documented API, which allows users to implement custom behavior.

The bundled modules perform common analysis tasks (e.g.,~clustering, filtering, equalization) and render plots in real-time (e.g.,~timeline, histograms), providing instant feedback that is particularly advantageous in experimental setting. Furthermore, the software can take data with Timepix2, Timepix3, embedded photomultiplier systems, and can orchestrate laboratory equipment (e.g.,~motorized stages, X-ray tubes). Thanks to ZeroMQ and memory-mapped file operations, data flow is implemented efficiently. This allowed Track Lab to be successfully deployed in many data-intensive applications over the past several years.

In the latest version, Track Lab is compatible with 7~popular operating systems (including 5~GNU/Linux distributions) and 2~CPU architectures. While binaries and documentation are publicly available for non-commercial use, source code is accessible upon request. Software contributions are welcome and encouraged.

\acknowledgments

The authors wish to thank Lukáš Meduna for his valuable contributions to Track Lab in initial stages of software development, Milan Beneš and Radu-Emanuel Mihai for their help with extensive on-premises testing. This work was carried out within Medipix Collaborations. The authors acknowledge funding from the Czech Science Foundation (GAČR) under grant number~GM23-04869M.

\bibliography{article}
\end{document}